\documentclass[graybox,vecphys]{svmult}

\usepackage{mathptmx}       
\usepackage{helvet}         
\usepackage{courier}        
\usepackage{type1cm}        
\usepackage{makeidx}         
\usepackage{graphicx}        
\usepackage{multicol}        
\usepackage[bottom]{footmisc}
\usepackage{amsmath}         

\newcommand{\bra}[1]{\langle#1|}
\newcommand{\ket}[1]{|#1\rangle}
\newcommand{\poly}{\text{poly}}
\newcommand{\id}{\mathbf{1}}

\makeindex             

\begin{document}

\title*{The computational complexity of density functional theory}
\author{James Daniel Whitfield, Norbert Schuch, Frank Verstraete}
\institute{James Daniel Whitfield \at VCQ, Universit\"at Vienna, Boltzmanngasse 5., Vienna, Austria \email{james.whitfield@univie.ac.at}
\and Norbert Schuch \at Institut f\"ur Quanteninformation, RWTH Aachen, D-52056 Aachen, Germany \email{schuch@physik.rwth-aachen.de}
\and Frank Verstraete  \at VCQ, Universit\"at Vienna, Boltzmanngasse 5., Vienna, Austria \email{frank.verstraete@univie.ac.at}}

\maketitle
\abstract{
	Density functional theory is a successful branch of numerical simulations of quantum systems.  While the foundations are rigorously defined, the universal functional must be approximated resulting in a `semi'-ab initio approach.  The search for improved functionals has resulted in hundreds of functionals and remains an active research area.  This chapter is concerned with understanding fundamental limitations of any algorithmic approach to approximating the universal functional.  The results based on Hamiltonian complexity presented here are largely based on \cite{Schuch09}.    In this chapter, we explain the computational complexity of DFT and any other approach to solving electronic structure Hamiltonians.  The proof relies on perturbative gadgets widely used in Hamiltonian complexity and we provide an introduction to these techniques using the Schrieffer-Wolff method.   Since the difficulty of this problem has been well appreciated before this formalization, practitioners have turned to a host approximate Hamiltonians.  By extending the results of \cite{Schuch09}, we show in DFT, although the introduction of an approximate potential leads to a non-interacting Hamiltonian, it remains, in the worst case, an NP-complete problem.  
}

\section{Introduction}

To tackle the limitations of simulating quantum systems a plethora of heuristics and approximate methods have emerged with density functional theory (DFT) at the forefront \cite{Parr89}. Other chapters in this book highlight recent advances in density functional theory, but this chapter will point out some of the ultimate limitations of any approach placed by computational complexity.  This chapter is largely an expansion of \cite{Schuch09} with additional results regarding Kohn-Sham DFT.

In this chapter, we will explore the difficulty of obtaining ground state energies of electronic Hamiltonians for the form:
\begin{equation}
	H_{elec}=T+W+V_{extern}+V_{mag}
\label{eq:elec}
\end{equation}
The effect of the magnetic field on the orbit of the electrons often requires a discussion of current density functional theory, but in this article the interest is only in the ground state and its energy. For the coupling of the electron orbitals and the magnetic field to contribute, currents are required but, as will be seen later, in the systems we utilize, the electrons are strongly localized in the ground state such that currents can only negligibly contribute.

In this chapter, we explain the worst case complexity of this Hamiltonian and show that any method, including DFT, capable of solving for the ground state energy of all Hamiltonians of the form given in eq.~\eqref{eq:elec} has solved a class of problems with implications far beyond electronic structure.  This parallels introduction of computational complexity into the context of spin systems.  Onsager's 1944 solution \cite{Onsager44} to the two-dimensional Ising model $H_{Ising}=-J\sum_{ij} Z_iZ_j$ generated many subsequent efforts to extend the method to three-dimensions, but in 1982 Barahona \cite{Barahona82} showed that obtaining efficient solutions to the three-dimensional Ising model was tantamount to solving an NP-complete problem.  The tremendous amount of research that is going into finding better functionals is facing a similar challenge but in some sense worse.  Here the problem is QMA-complete which subsumes NP-completeness.  The definitions and motivations behind the terms QMA and NP will be provided later in this chapter.

The complexity of  DFT as originally formulated can be dealt with using extensions of Hamiltonian complexity to the electronic structure problems captured by eq.~\eqref{eq:elec}, but upon introducing an approximate functional and retreating to the Kohn-Sham formalism, there are important modifications to be considered.  In this chapter, we  extend unpublished results for Hartree-Fock to Kohn-Sham DFT and show that the self-consistent field method widely employed for DFT leads to problems within the NP complexity class.

In the next section, we will discuss background material, namely, density
functional theory, Hamiltonian complexity, and perturbation theory. In the
following section, we explain a chain of reductions that allows us to reduce the Hamiltonian presented in eq.~\eqref{eq:elec} to other problem that have previously been shown to be difficult.  Next, we explore consequences of these reductions and some limitations to this approach.  Finally, we discuss computational complexity of Kohn-Sham DFT in Section \ref{sec:KS-DFT} before offering concluding remarks in Section \ref{sec:final}.

\section{Background}\label{sec:background}
In this section, we present background material to make the key ideas accessible.
\subsection{Density functional theory}
Density functional theory, as explained in several other chapters, is predicated on the use of the one-particle probability density, $n(r)$, as the fundamental variable in place of the $N$-body wave function, $\Psi$. This idea is founded upon the Hohenberg-Kohn theorems \cite{Hohenberg64}.  The first theorem proves that the density is uniquely determined by the external potential for systems with non-degenerate ground states.  The second theorem provides a variational principle for the density.

In the same paper, the idea of the universal functional was introduced defined as
\begin{equation}
	F[n(r)]=\bra{\Psi_0} (T+W) \ket{\Psi_0}.
	\label{eq:F}
\end{equation}
with $\Psi_0$ the ground state wave function from which $n(r)$ arises. Issues regarding $v$-representability can be dealt with using the Levy minimization procedure \cite{Levy79}.  If this functional could be approximated then the ground state energy could be easily extracted since the minimization of the energy can be done efficiently.  This follows as the set of $N$-body density matrices, 
$\rho$, form a convex set\footnote{If $\rho_i$ are density matrices then $\rho=\sum p_i\rho_i$ is also whenever $p_i>0$ and $\sum p_i=1$.}.  While the Hohenberg-Kohn theorems allow the probability density to be used as the 
basic variable, this work alone did not provide sufficient reason for doing so.

It was the Kohn-Sham (KS) construction \cite{Kohn65} introduced the following year that allowed practical DFT technologies emerge from the algorithm developed for Hartree-Fock: the self consistent field (SCF) method.  In the KS construction, the interacting system is replaced by a non-interacting (free-fermion) system with a different single electron potential that is designed to reproduce (in the non-interacting system) the same density as the interacting system.  

\subsection{Hamiltonian complexity}
One the key contributions of quantum information theory has been Hamiltonian complexity \cite{Osborne12}.  This extends the work of Barahona \cite{Barahona82} discussed in the introduction from Ising type Hamiltonians to quantum Hamiltonians with off-diagonal couplings and allows fruitful generalizations of NP-completeness to the quantum setting. 

In Hamiltonian complexity, the amount of computation required to verify the energy of the ground state up to a pre-specified accuracy dictates the complexity of the Hamiltonian.  If, on one hand, a classical computer, can verify the energy of a proposed state in an amount of time that is a polynomial of the input size then the Hamiltonian is considered NP-hard.  On the other hand, if it requires a quantum computer to verify the state's energy in polynomial time, then the problem is called QMA-hard.  QMA stands for `quantum Merlin-Arthur' and NP is `non-deterministic polynomial.' If the given Hamiltonian has enough flexibility to encode any other problem in the NP complexity class, then the Hamiltonian is NP-complete.  Similarly, if the Hamiltonian can encode any other QMA problem, then the problem is QMA-complete.

The first QMA-complete Hamiltonian was introduced by Kitaev
\cite{Kitaev02} building on ideas of Feynman\cite{Feynman82}.  The
Hamiltonian construction represents an updated version of the Cook-Levin
construction which shows any NP-complete problem can be embedded into the
Boolean satisfiability problem \cite{Sipser97}. The Hamiltonian is constructed such that any problem that can be verified by a quantum computer can be embedded into the ground state of this Hamiltonian.  That is, if, for some problem, there is some verification scheme with a quantum computer that can check that the input state is correct quickly (i.e. in a time that scales only polynomially with the input size, as contrasted with exponentially scaling verification procedures)  then this problem can be embedded in to the following Hamiltonian and its associated ground state problem.

The clock construction is designed such that the ground state of the Hamiltonian encodes the history of a quantum computation.
\begin{eqnarray}
H_{QMA5}&=&H_{init} + H_{trans.}+ H_{final}\\
&=&\ket{\psi_{t=0}}\bra{\psi_{t=0}}\otimes\ket{t=0}\bra{t=0}\nonumber\\
&& +\left( \sum_{t=1}^T U_t\otimes\ket{t+1}\bra{t} +U_t^\dag\otimes\ket{t}\bra{t+1}\right)\nonumber\\
&& + \ket{\psi_{t=T}}\bra{\psi_{t=T}}\otimes\ket{t=T}\bra{t=T}
\label{eq:H5}
\end{eqnarray}
The ground state of this Hamiltonian is special and is given by
\begin{equation}
\ket{\Psi}=\sum_{t=0}^T\ket{\psi_t}\otimes\ket{t}=\sum_{t=0}^TU_t...U_2U_1\ket{\psi_0}\otimes\ket{t}
\label{eq:hist}
\end{equation}

The first register of \eqref{eq:hist}, $\psi_t$, is called the computation register and stores the state of the verification procedure at any given time.  The second register, $t$, is called the clock register. This register is used to keep track of the computation's progress and is composed of $T$ bits where $T$ is the length of the circuit.  Valid clock register states are required to have a single domain wall between the zeros and the ones.
For example,
  \[
\ket{t=3}=\ket{1110000...0}
\]
The location of the domain wall is correlated with the computation register such that after $m$ gates have been applied in the computation register, the domain wall is between sites $m$ and $m+1$ representing $\ket{t=m}$.

The requirement that the state contains five-spin interactions comes from the two-spin gates required for universal quantum computation \cite{Nielsen00} and the three-spin check needed to evaluate the clock register's domain wall location. The circuit that is embedded into the ground state corresponds to the verification procedure of the QMA problem being mapped to the $H_{QMA5}$ problem.

Through the use of perturbation theory this Hamiltonian can be reduced to a two-spin Hamiltonian \cite{Kempe06}. 
\begin{equation}
	H_{QMA2}=\sum_{ij} J_{ij}A_iB_j
	\label{eq:qma2}
\end{equation}
with $A$ and $B$ one of the Pauli matrices: 
\begin{align}
&X=\left[\begin{array}{cc} & 1\\1& \end{array} \right]&
&Y=\left[\begin{array}{cc} & -i\\i& \end{array} \right]&
&Z=\left[\begin{array}{cc}1 & \\&-1 \end{array} \right]&
	\end{align}
Note for QMA-completeness, not all Pauli couplings are needed~\cite{Biamonte08} and spatial locality for the couplings can be imposed~\cite{Oliveira08}. The perturbative techniques required to convert the five-spin Hamiltonian to the two-spin Hamiltonian, play an important role throughout the remainder of the paper.  For that reason, we will present the basic techniques utilized later in this section.  But first, we define our notions of precision. 

\subsection{Accuracy}

For Hamiltonian problems such as Ising where the spectrum is discrete
obtaining the exact value of the ground state energy is possible, but in the general setting verifying the ground state energy cannot be done exactly since the spectrum is real valued and possibly irrational.  Instead the energy should be defined up to some precision. In this section, we will discuss what precision is necessary for the problem to be QMA and explain why it coincides with the typical setting that one is interested in. 

To illustrate the issues at hand, consider the problem of obtaining the energy of some Hamiltonian $H$ using an algorithm that returns values at a fixed precision of $\delta=0.1$ energy units.  If you are interested in the energy at, say, $0.001$ energy units, you could just multiply $H$ by 100 and utilize the same algorithm.  By rescaling the Hamiltonian, one could obtain the energy to arbitrary accuracy.

Of course, this is not plausible, otherwise there would be no need to bound the accuracy at all.  In reality, the accuracy is specified relative to the size of the input.  Continuing with the illustration above, consider that the Hamiltonian entries are specified by single precision binary floating point variables.  By rescaling the Hamiltonian by some large factor, will require double, quadruple, or larger precision floating-point variables to preserve the integrity of the input.  Thus, it becomes clear that the algorithm which was designed to return a fixed precision of $\delta$ is not appropriate for the rescaled Hamiltonian without some modifications. 

In the context of Hamiltonian complexity, this rescaling issue is handled by forcing the precision to scale as an inverse polynomial in the number of unit strength interaction terms of the Hamiltonian. More precisely, given a Hamiltonian
\[H=\sum_i^n h_i\]
where in the operator norm, $|h_i|\leq 1$, then for the problem to be in QMA the precision must be $\delta\leq \poly(n)$. If one or more of the interaction terms has norm greater than 1, say $J$, then replace $H$ with a rescaled Hamiltonian, $H'=H/J$.

We can show why this scaling is enforced and show that it is well motivated by returning to the illustration. As the system size increases, we naturally expect the energy to grow extensively with the system size.  Suppose, as before, we have access to an MEASURE0 algorithm that measures energies only between $\pm E_0$ where $E_0$ could be, for instance, 1 eV. Suppose the problem of interest has an energy scale that is $E_s=\poly(s)E_0$ and the desired precision $\delta_s$ is fixed at some numerical value, e.g. $0.001$ eV. To utilize MEASURE0 for this problem, a simple calculation 
\begin{eqnarray}
	E_s\pm \delta_s = \frac{E_s}{E_0} E_0\pm \delta_s = \frac{E_s}{E_0} (E_0\pm \delta) 
\end{eqnarray}
implies that $\delta=\delta_s/\poly(s)$.  Since $\delta_s$ is independent of $s$, $\delta$ must scale as an inverse polynomial in $s$. Whether it is possible to utilize MEASURE0 with a fixed accuracy to address the $E_s\pm \delta_s$ problem relates to the modern study of approximatability in computer science and probablistically checkable proofs (PCP).

The PCP theorem provides insights into the ability to verify proofs probabilistically and has spawned many important results \cite{Arora1998a,Arora1998b,Arora09}. The relevant consequence of the theorem in this context, is the implied hardness of approximation for some NP-complete problems.  For instance, in max-3-SAT, for an arbitrary set of clauses containing the conjunction of three variables (or their negations), one can guess randomly that 7/8 of the clauses are satisfiable based on simple probability arguments.  However, deciding if more than 7/8 are satisfiable is known to be an NP-complete problem as a consequence of the PCP theorem. It is an active research area to see if such hardness of approximation extends to the quantum regime discussed in this article.

\subsection{Perturbation theory in Hamiltonian complexity}  

The perturbation theory used in this article is based on the Schrieffer-Wolff transformation. For a modern introduction see \cite{Bravyi11}. In our situation, the unperturbed Hamiltonian, $H$, has a spectral gap $\Delta$ and we will divide the system into a low and high energy sector.  A perturbation, $\varepsilon V$, is then included which couples the two sectors.
\begin{equation}
H=\left[\begin{array}{cc}H_0 &  \\  & H_1\end{array}\right] +\varepsilon\left[\begin{array}{cc}V_0 & V_{01} \\ V_{10} & V_{1}\end{array}\right] 
\end{equation}
This perturbation will introduce effective interactions in the low energy subspace enabling a target Hamiltonian incorporating these new interactions to be enacted up to some order in an $\varepsilon$ and $1/\Delta$ expansion. The Schrieffer-Wolff method used in degenerate perturbation theory relies on a unitary transform, $U=\exp(S)$, that maximally separates the high energy and low energy sectors of the Hamiltonian. 

To obtain the expansion to second order, we will construct $S$ using an expansion in $\varepsilon$,
\begin{equation}
	S=\sum S_k \varepsilon^k=\sum \left(\begin{array}{cc} &X_k\\-X^\dag_k\end{array}\right)\varepsilon^k,
\end{equation}
and then apply the BCH formula to the transformed Hamiltonian
\begin{eqnarray}
	&&e^S H_{tot} e^{-S}= \left(\begin{array}{cc}
		H_{eff} & \\& H_{high}
	\end{array}\right) +O(\varepsilon^3)\\
	&=&H+\varepsilon(V+[S_1,H])+\varepsilon^2\left([S_2,H]+[S_1,V]+\frac12 [S_1,[S_1,H]]\right)+O(\varepsilon^3)
\end{eqnarray}
At first order in $\varepsilon$, the off-diagonal block should vanish leading to $0=(V+[S_1,H])_{01}=V_{01}+X_1H_1-H_0X_1$. Solving for $X_1$, we have $X_1=-V_{01}H_1^{-1}+H_0X_1H_1^{-1}$. Dropping the last term yields the approximation $X_1=-V_{01}H_1^{-1}$ which is correct to leading order in $1/\Delta$. 

Doing the same for the next order in $\varepsilon$ leads to $$0= \left(X_2H_1-H_0X_2+X_1V_1-V_0X_1\right)_{01}+ H_0V_{01}H_1^{-1}$$ where the last term is the correction from the first order approximation and the double commutator $[S_1,[S_1,H]]$ did not contribute since it is quadratic in $S$. Rearranging, we get
\[
	X_2=-H_0V_{01}H_1^{-2}+V_{01}H_1^{-1}V_1 H_1^{-1} -V_0V_{01}H_1^{-2}
	\]
where, as before, we dropped a term from the right hand side containing $X_2$ which can only contribute at higher orders.

With the expansion for $S$, it is straightforward, albeit tedious, to compute 
\begin{equation}
	H_{eff}=H_0+V_0-V_{01}H_1^{-1}V_{10}+O(\epsilon^3/\Delta^2).
\end{equation}


In the next section, the perturbative gadgets will be used to obtain effective low energy descriptions from a restricted type of Hamiltonian.  Let us present the general structure, beginning with a system composed of $N$ spin systems.  The unperturbed Hamiltonian is
\begin{equation}
	H=\Delta \; \id_{N-1}\otimes \ket{h}\bra{h}=
	\Delta\left(\begin{array}{c|c}
		\phantom{\id_{N-1}}& \\\hline &  \id_{N-1}\\
	\end{array} \right)
\end{equation}
where the matrix is written in the energy splitting basis $\{\ket{h},\ket{l}\}$. We can parameterize the energy splitting basis using rotation matrix $M(\theta,\varphi)=\left(\begin{smallmatrix}\cos\theta&-\sin\theta e^{-i\varphi}\\\sin\theta e^{i\varphi}&\cos\theta\end{smallmatrix}\right)$ such that $M\ket{1}=\ket{h}$. The perturbation acting on spin $N$ which will generate desired interactions is given by
\begin{eqnarray}
	V&=& \epsilon\sum_{ijk}v_{ijk} 
	\;M^\dag
	\left(
	\begin{array}{c|c}
		C_k & A_i-iB_j\\
		\hline
		A_i+iB_j & -C_k
	\end{array}\right) M\\
	&=& \epsilon \sum_{ijk}v_{ijk}\; M^\dag (A_i\otimes X + B_j\otimes Y+C_k\otimes Z)M
	\label{eq:pert}
\end{eqnarray}
Here $A_i$, $B_j$, and $C_k$ are arbitrary Pauli matrices acting on spins $i$, $j$, and $k$, respectively. Unlabeled Pauli matrices act on the $N$th spin.
To compute the effective Hamiltonian, we need the following quantities which are easily obtained
\begin{subequations}\label{eq:offdiag}
	\begin{eqnarray}
		\bra{l}X\ket{h}=\bra{0}M^\dag X M\ket{1}&=&\cos^2\theta-\sin^2\theta e^{-2i\varphi}\\
	\bra{l}Y\ket{h}=\bra{0}M^\dag Y M\ket{1}&=&-i(\cos^2\theta+\sin^2\theta e^{-2i\varphi})\\
	\bra{l}Z\ket{h}=\bra{0}M^\dag Z M\ket{1}&=&-\sin2\theta e^{-i\varphi}
	\end{eqnarray}
\end{subequations}

Upon expanding, we will get terms that are one-local at both first and second order in $\varepsilon$ and will be grouped into $H_{loc}$. It is important to note that these terms are within the low energy effective space of spin $N$ and thus do not cause excitations within the splitting basis. The remaining terms which come from $V_{01}V_{10}$ contain parameterized couplings which can be computed by expanding \eqref{eq:pert} using \eqref{eq:offdiag}. The end result is
\begin{eqnarray}
	H_{eff}&=& H_{0}+\varepsilon V_0-\varepsilon^2\frac{V_{01}V_{10}}{\Delta}\nonumber\\
	&=&H_{loc}+ \frac{\varepsilon^2}{\Delta}\left(\sin^22\theta\,\sin2\varphi A_iB_j-\cos\varphi\,\sin4\theta\; A_iC_k+\sin4\theta\,\sin\varphi\; B_jC_k\right)
	\phantom{seee}
	\label{eq:heff}
\end{eqnarray}

%
%

\section{The QMA difficulty of $H_{hubbard}$}\label{sec:QMA}
In this section we demonstrate that the Hamiltonian \eqref{eq:elec} is QMA-complete.  The proof strategy is based on a long standing idea of computer science: that of reductions.  Shortly after the concept of NP-completeness was introduced, 21 additional problems  were shown to be NP-complete \cite{Karp72}.  The strategy that was used, and that is employed here, is to show that problem $A$ reduces to another problem $B$ such that if you could solve $B$ then with a little more effort you could solve $A$.  We denote this relationship with $A\leq B$ with the understanding that $A$ and $B$ represent classes of problems. 

In the quantum case, one can play a similar game. 
In the context of Hamiltonian complexity, the idea of reductions is akin 
to the embedding of ground state problem for Hamiltonian $H_A$ into instances of the ground state problem for Hamiltonian $H_B$.   Note that the mapping used to show $H_A\leq H_B$ can, and often will, require solving $H_B$ on a larger system size and requiring more interactions than that of $H_A$.  So long as the system size and the system resources required scale as a polynomial of the system size, then this mapping is considered efficient. 

The remainder of this section is used to present a series of reductions that ultimately demonstrate that $H_{QMA2}$ can be embedded into instances of $H_{hubbard}$.  Since the Hubbard model is a phenomenological description of the electronic Hamiltonian, this implies that algorithms solving $H_{elec}$ could also be used to solve $H_{QMA2}$.  The Hamiltonians used in the reduction are the Heisenberg model and the Hubbard model which are also interesting in their own right.

\subsection{Proof}
The proof proceeds by demonstrating that the known  QMA-complete Hamiltonian $H_{QMA2}$ can be embedded into the Heisenberg model, that the Heisenberg model can be embedded into the Hamiltonian of the Hubbard model, and finally that the Hubbard model arises from an electronic Hamiltonian as in \eqref{eq:elec}.

\subsection{$QMA2 \leq Heisenberg$}
The goal of this subsection is to demonstrate that finding the ground states of the Heisenberg Hamiltonian is QMA-hard. The Hamiltonian is defined by
\begin{equation}
	H_{heisenberg}=J\sum_{ij} X_iX_j+Y_iY_j+Z_iZ_j + \sum_i\mathbf{b}_i\cdot\mathbf{S}_i
	\label{eq:heis}
\end{equation}
where $\mathbf{b}=[b_x\;b_y\; b_z]$  and $\mathbf{S}=[X\;Y\;Z]$ with $X,Y,Z$ the Pauli matrices previously defined. This is done using the perturbative gadgets introduced earlier.  To complete the embedding,  two types of gadgets are necessary.  The first is used to embed arbitrary terms in \eqref{eq:qma2} into a Hamiltonian of with a standard format and the remaining gadgets remove undesired interaction without changing the low energy subspace.  

The first gadget is designed to embed arbitrary couplings $J_{ij}A_i\otimes B_j$ into a Hamiltonian with uniform strength couplings between equivalent Pauli matrices.  That is $J_{ij}A_iB_j+H_{loc}\leq \lambda (A_i\otimes A_N+B_j\otimes B_N) +H_{loc}$ with $\lambda$ site independent.  Using $V=\lambda(A_i\otimes A_N+B_j\otimes B_N)$ and the parameterization of $H_{eff}$ in \eqref{eq:heff}, we can readily identify how to create couplings by selecting $\theta=\pi/8$ if either $A$ or $B$ is the Pauli $Z$ matrix and $\theta=\pi/4$ otherwise. Then $\varphi$ provides a tunable parameter to adjust the coupling strength. In the case, that $A$ and $B$ are the same, the Hamiltonian cannot be directly embedded since the perturbative corrections would always be of fixed sign 
(because $V_{01}V_{10}=|\bra{l}M^\dag P M\ket{h}|^2A_iB_j>0$). Instead, the coupling is first decomposed into fixed strength couplings with different Pauli matrices and then decomposed as before. The strength of the perturbation, $\lambda$ should be such that $\lambda^2/\Delta=1$ allowing $\varphi$ alone to dictate the coupling.  

The second type of gadgets are those used to enable more complex Hamiltonians contain undesired interaction to embedded simpler ones.  This is done by pushing the unwanted interactions out of the effective Hamiltonian using a strong field applied in the eigenbasis of the interaction so as to freeze all low energy states into an eigenstate of the interaction. This is clear upon returning to \eqref{eq:offdiag}.  For any Pauli matrix, $P$, when the splitting basis is eigenbasis of $P$ then the coupling term is $\bra{l} M^\dag P M\ket{h}=\bra{p_0}P\ket{p_1}=0$.  

With the description of the two types of gadgets necessary, let us specify the actual chain of reductions (ignoring the single spin terms):\
\begin{eqnarray}
J_{ij}A_iB_j +O\left(\frac{\lambda_1^3}{\Delta_1^2}\right)&\leq & \lambda_1 (A_iA_1 + B_jB_1)\label{eq:G1}\\
\lambda_1 A_mA_n +O\left(\frac{\lambda_2^3}{\Delta_2^2}\right)&\leq & \lambda_2 [A_mA_2 +B_mB_2\;+ A_nA_2+B_nB_2]\label{eq:G2}\\
\lambda_2(A_a A_b+B_aB_b)+O\left	(\frac{\lambda_3^3} {\Delta_3^2 } \right) &\leq & \lambda_3(\mathbf{S}_a\cdot\mathbf{S}_3+\mathbf{S}_b\cdot\mathbf{S}_3)\label{eq:G3}
\end{eqnarray}
At this point, let us remark that the error listed is for each gadget and when summing all terms of the Hamiltonian, the error terms are summed (for an upper bound).  After the three gadget layers there will be eight Heisenberg couplings for each coupling in $H_{QMA2}$.  The first layer is accomplished with the first gadget discussed and remaining gadgets are of the second kind.  Remember that if in $H_{QMA2}$, $A_i=B_j$ then one must include an additional gadget layer to allow for positive and negative couplings. 

The final consideration is the appropriate energy scales of each gadget.  For each gadget we have three rules to ensure that the gadgets do not `cross-talk' and the perturbation expansion used remains reliable. 
\begin{enumerate}
\item $\Delta_i\gg \lambda_i$. The splitting field should be the dominate energy scale.
\item $\lambda_i\gg\Delta_{i-1}$. The previous gadget should be on a lower energy scale.
\item $\lambda_{i-1}=\lambda_i^2/\Delta_i$. This follows directly from \eqref{eq:heff}.
\end{enumerate}
For a precision of $\delta\leq 1/\text{poly}(N)$, the splitting and the coupling must be such that $\lambda_i^3/\Delta_i^2\ll\delta$.  Recall in the first gadget that $\lambda_1^2=\Delta_1$ to allow $\varphi$ to control the coupling strength.  

Following these rules, one arrives at coupling and splitting strengths that scale extremely poorly with the number of sites but still only of polynomial order and hence efficient from a theoretical perspective.  
	
\subsection{ $Heisenberg\leq Hubbard$}
The second embedding of the Heisenberg model into the Hubbard model at half-filling is well known.   
The Hubbard model has the Hamiltonian
\begin{equation}
	H_{hubbard}=t\sum_{ij}\sum_{\sigma\in\{\downarrow,\uparrow\}} a_{i\sigma}^\dag a_{j\sigma} + U\sum_i a_{i\uparrow}^\dag a_{i\uparrow}a_{i\downarrow}^\dag a_{i\downarrow}+\sum_i \mathbf{b}_i\cdot \mathbf{S}_i
\end{equation}
with creation and annihilator operators satisfying fermionic anti-commutation relations: $a_{i\sigma}a^\dag_{j\sigma'}+a^\dag_{j\sigma'}a_{i\sigma}=\delta_{ij}\delta_{\sigma\sigma'}$ and $a_{i\sigma}a_{j\sigma'}+a_{j\sigma'}a_{i\sigma}=0$.  The vector $S$ is defined as as before with the understanding that Pauli matrices are expressed in terms of the creation/annihilation operators as $P_i= P_{\sigma\sigma'}a_{i\sigma}^\dag a_{i\sigma'}$,

As before, we will use second order perturbation theory with $H=U\sum_i n_{i\uparrow}n_{i\downarrow}$ as the unperturbed Hamiltonian and $V=t\sum a_{i\sigma}^\dag a_{j\sigma}$ serving as a perturbation.  At half filling, the perturbation can only cause excitations away from the low energy space, hence $V_{0}=0$.  However, at second order in the interaction, the electrons can interact in the low energy subspace and this is captured by the Heisenberg Hamiltonian.  To prove this statement consider the off-diagonal block of $V$ with respect to $H$,
\begin{eqnarray}
	V_{10}= t\sum_{ij}\sum_\sigma P_H(i,\sigma) a^\dag_{i\sigma}a_{j\sigma}P_L(j,\sigma)
\end{eqnarray}
where the local projector onto the low energy subspace is given by $P_L(i,\sigma)=1-n_{i\bar{\sigma}}$ with $\bar\sigma$ the opposite of $\sigma$. The high energy projector is $P_H=\id-P_L$. 

Physically, the electron from  one site, say $i$, hops to another site, say $j$, through the $V_{10}$ term and, to remain in the low energy subspace, it must return to its original position through $V_{01}$ but there are two ways to return to the low energy subspace providing the desired interaction terms. Considering the process yields  
\begin{eqnarray}
	H_{eff}&=& \frac{t^2}{U} \left( V_{01}V_{10} \right)\\
	&=& \frac{t^2}{U}\sum_{\sigma,\tau} P_L(i,\sigma) a^\dag_{i\sigma}a_{j\sigma}P_H(j,\sigma)P_H(j,\tau)a^\dag_{j\tau}a_{i\tau}P_L(i,\tau)\\
	&=& \frac{t^2}{U}\sum_{\sigma,\tau} a^\dag_{i\sigma}a_{j\sigma}a^\dag_{j\tau}a_{i\tau}\\
	&=& \frac{t^2}{U}\sum_{\sigma,\tau}\left(\delta_{\sigma\tau}a_{i\sigma}^\dag a_{i\sigma}-a^\dag_{i\sigma}a^\dag_{j\tau}a_{j\sigma}a_{i\tau}  \right)
	\label{eq:exchange}
\end{eqnarray}
  Using a matrix representation quickly illustrates that this gives rise to the Heisenberg coupling:
\begin{equation}
	\sum_{\sigma\tau}a^\dag_{i\sigma}a^\dag_{j\tau}a_{j\sigma}a_{i\tau}=
{\let\quad\qquad
	\bordermatrix{&a_{j\alpha}a_{i\alpha}&a_{j\alpha}a_{i\beta}&a_{j\beta}a_{i\alpha} &a_{j\beta}a_{i\beta}\cr\\[1em]
		(a_{j\alpha}a_{i\alpha})^\dag&1 &    &  & \cr\\[1em]
                (a_{j\alpha}a_{i\beta})^\dag&   &   &1  & \cr\\[1em]
               (a_{j\beta}a_{i\alpha})^\dag &  &1 &     & \cr
                (a_{j\beta}a_{i\beta})^\dag&   &     & &1 }
	}
\end{equation}
Similarly expanding $\mathbf{S}_i\cdot\mathbf{S}_j=X_iX_j+Y_iY_j+Z_iZ_j$ yields
\[
	\mathbf{S}_i\cdot\mathbf{S}_j=
\begin{pmatrix}
	1& & &\\
	&-1&\phantom{-}\phantom{-}2&\\
	&\phantom{-}2&\phantom{-}-1&\\
	& & &\phantom{-}1
\end{pmatrix}
\]
from which it should be clear that 
\begin{equation}
	H_{eff}= H_{loc}-\frac{t^2}{2U}\mathbf{S}_i.\mathbf{S}_j+\frac12 \id +O\left(\frac{t^3}{U^2}\right)
\end{equation}

The choice of parameters for $t$ and $U$ are restricted by the same rules as before:  (1) $U\gg t$, (2) $t\gg \Delta_3$ and (3) $\lambda_3=t^2/2U$ with  $\Delta_3$ and $\lambda_3$ referring to the gadget in \eqref{eq:G3}.


\section{Consequences of reductions}
From the series of reductions, we see that within the Hamiltonian, $H_{hubbard}$, we can embed instances of $H_{QMA2}$.  This has certain implications for density functional theory which we now discuss. 

For mixed states, the optimization over possible input states can be done efficiently since the set of $N$-electron density matrices forms a convex set.  In this case, evaluating the universal functional would allow us to obtain the energy of Hamiltonian $H_{elec}$ realizing the Hubbard model efficiently.  Of course, due to the second Hohenberg-Kohn theorem, as discussed above, the density corresponding to the lowest energy will correspond to $\rho=\ket{\Psi_0}\bra{\Psi_0}$ for non-degenerate ground state $\Psi_0$.

Since the final embedding of the Heisenberg Hamiltonian into the Hubbard Hamiltonian requires the electronic Hamiltonian to be at half filling, it is in an insulating phase where currents are not present.  This rationalizes the lack of current density functional theory which would ordinarily be needed when dealing with magnetic fields.

\section{Kohn-Sham DFT}\label{sec:KS-DFT}
Until now we have focused on the complexity of DFT as was originally
formulated using the universal function for the kinetic energy and the
Coulomb interaction.  While this is an interesting construction from a
mathematical physics point of view, it was really the introduction of the KS formulation that allowed DFT to become such a successful numerical technique.  Like its precursor Hartree-Fock \cite{Szabo96}, Kohn-Sham is based on a single determinant describing a non-interacting system.  Hartree-Fock's success, in many ways, can be attributed to suitably approximating the kinetic energy operator and Kohn-Sham builds upon this success by reformulating the HF method in terms of the functional derivative of the universal functional to obtain the scalar exchange-correlation potential. 

The self consistent field method developed for Hartree-Fock directly applies to KS-DFT where the non-linear eigenvalue problem is modified by the presence of an approximate exchange-correlation potential.  In both situations, the non-interacting Hamiltonian is formed based on the previous eigenvalues and eigen-orbitals, diagonalized for new eigenvalues and eigen-orbitals, and then recomputed until self-consistency.  This algorithm can be formulated as an optimization problem if we consider 
\begin{equation}
	E=\min_{\Psi\in SD_1} \bra{\Psi}H\ket{\Psi}
	\label{eq:mean}
\end{equation}
with $H=T+V+W+V_{xxc}$ where $SD_1$ is the set of all single Slater determinants, $V_{xxc}=V_{xc}-E_x^{HF}$ is the modified exchange correlation potential, $V_{xc}$ is obtained as functional derivative of an approximate exchange correlation functional and $E_x^{HF}$ is the exact exchange from Hartree-Fock. The Hartree-Fock optimization procedure is identically formulated except without $V_{xxc}$ so the proof given below equally applies to Hartree-Fock. 
Now we will show that optimization problems of the form \eqref{eq:mean} are in NP-complete complexity class. This result is not surprising in light of the many research articles dedicated to accelerating and ensuring convergence of SCF methods e.g. \cite{Kudin07}.

We consider the equation~\eqref{eq:mean} in second quantization with Hamiltonian
\begin{equation}
\label{hf-ham}
 H=\sum^{M^2}_{ij} h_{ij}a^\dagger_ia_j+
\frac12 \sum^{M^4}_{ijkl}\; h_{ijkl}\;a^\dagger_ia^\dagger_ja_ka_l.
\end{equation}
with a number of orbitals $M\ge N$. In second quantization, the set of single Slater determinants are
defined by $SD_1= \{b^\dagger_N\cdots b^\dagger_1\ket\Omega\}$ with
$b_i=\sum u_{ij} a_j$  and  $\ket\Omega$ as the vacuum. 

Consider the problem of computing the energy according to \eqref{eq:mean} up to precision $\delta<1/\poly(N)$. We proceed as before by reducing this problem to another using perturbative embeddings.
Note that the energy can be verified efficiently classically using the Slater-Condon rules~\cite{Szabo96}. 

We show that this problem can embed  Ising spin glasses which are
known to be NP-hard~\cite{Barahona82}: Given an $L\times L\times 2$ lattice
of two-level spins $S_i=\pm 1$ with a nearest neighbor
Ising coupling $H_{Ising}=\sum J_{ij}S_iS_j$, $J_{ij}\in\{0,-1,1\}$,
determine whether the ground state energy is the minimum one allowed by
the individual $J_{ij}$'s or not. 

The technique used for the proof is almost identical to the embedding of the Heisenberg 
model into the low energy sector of the Hubbard model.  Again we consider the system at 
half filling where $M=2N$ and consider unperturbed Hamiltonian
$H=U\sum_i a_{i\uparrow}^\dag a_{i\downarrow}^\dag a_{i\downarrow}a_{i\uparrow}$.  
Just as before, we convert the spin operator $Z$ to fermionic modes via 
$a_\sigma^\dag Z_{\sigma\sigma'}a_{\sigma'}$ giving $V=\sum_{ij}J_{ij}Z_iZ_j=J_{ij}\sum_{p,q=0,1}(-1)^{p+q} n_{2i+p}n_{2j+q}$.  The resulting product is quadratic and of form~\eqref{hf-ham}.

In the effective Hamiltonian, the first order correction, $V_0$, gives the Ising energies and errors arise only at second order.  Since we are embedding the Ising Hamiltonian, there are $O(N^2)$ terms in the summation and the maximum absolute value of each term is unity. Hence, $U$ can be estimated as $O(N^2)$. So long as $\delta<O(N^{-2})$, the first order corrections occurring at order $U^{-1}$ can be distinguished and the ground state Ising energies can be recovered.

Since the ground state of the system is a classical spin state,
it can be expressed as a Hartree-Fock state where $b_i=a_{i\uparrow}$ or
$b_i=a_{i\downarrow}$, respectively, and since the classical Hamiltonian has a
constant gap while perturbations from the penalized
subspace are at most $O(1/U^2)$, a polynomial
accuracy is sufficient to make the problem NP-hard.

\section{Conclusions}\label{sec:final}
	
In this chapter, we have reexamined the difficulties facing density functional theory by examining the complexity of the electronic structure Hamiltonian with local magnetic fields.  There are limits to the applicability of the results since the problem is encoded into the local magnetic fields.  The existence of purely electronic potential that can be rigorously shown as NP or QMA-complete remains unproven.  Of course, we believe such a construction exists and perhaps this chapter will help light the path forward.

Finally, we stress that no matter the complexity of the problems at hand, there can and has been tremendous numerical triumphs. NP-completeness or QMA-completeness is only a worst case analysis and does not probe a particular ensemble of interesting problems nor ask where the difficult problems lay. Thus, the worst case complexity only informs one that there exists fundamental boundaries but does not give any indication of the distance from the wall. 

\paragraph{\textbf{Acknowledgements}}
We would like to thank S. Andergassen for helpful comments on the manuscript and JDW thanks the VCQ Postdoctoral Fellowship.

\end{document}